\documentclass[a4paper,11pt]{article}
\usepackage{jheppub} 
\usepackage{lineno}
\usepackage{graphicx}
\usepackage{pgfplots}
\pgfplotsset{compat=1.18}
\usepackage{caption}
\usepackage{subcaption}
\usepackage{float}
\usepackage{empheq}
\usepackage{amsmath}
\usepackage{amssymb}
\usepackage{mathtools}
\usepackage{physics}
\usepackage{lipsum}

\title{\boldmath Bounds on CFT correlations from the thermal partition function}

\author{Fikret Ceyhan, Thomas Faulkner}
\affiliation{Department of Physics, University of Illinois, Urbana-Champaign,\\
1110 W. Green St., Urbana IL 61801, USA.}

\emailAdd{fceyhan2@illinois.edu}
\emailAdd{tomf@illinois.edu}

\abstract{We discuss upper bounds on the mutual information for disjoint spherical regions of the CFT vacuum. To prove our bounds, we utilize the modular nuclearity condition, which is in turn related to finiteness of the thermal partition function of the CFT. Our bounds are satisfied by the conjectured geometric duals of these correlation measures in AdS/CFT, where they link the Hawking-Page phase transition with the entanglement wedge phase transition. We use these results to conjecture a new boundary theory quantity that computes the minimal distance between two general entanglement wedges.}

\usepackage[compatibility=false]{caption}
\begin{document}
\maketitle
\flushbottom

\label{sec:intro}

\section{Introduction}
\label{sect:1}

\noindent
Studying the algebraic inclusion of space-like regions described by nested causal diamonds captures
information on correlations in the QFT vacuum \cite{Pasquale_Calabrese_2004}. More surprisingly, it also captures aspects of excited states, such as the thermal state, essentially because the density matrix of a localized region in a relativistic theory
looks thermal.  
Tomita-Takesaki modular theory of von Neumann algebras provides the rigorous mathematical toolkit for linking the geometric properties of subregions with the thermodynamic properties of the underlying theory. In this paper, we will focus on a specific tool called modular nuclearity \cite{2007CMaPh.270..267B}.
We will study this aspect of modular theory using the $AdS/CFT$ correspondence \cite{Maldacena_1999, witten1998antisitterspaceholography, Gubser_1998} where the link between algebras, correlations, and thermodynamics is particularly strong \cite{Headrick_2014}, as captured by the link between boundary regions, extremal surfaces, and the black hole thermodynamics of generalized horizons. 


In recent years, modular theory has played an indispensable role in gaining a deeper understanding of this duality.
One such example revisited the Hawking-Page temperature $T_{HP}$ from this algebraic perspective.
It was recently argued that the operator algebra associated with one-half of the thermofield double Hilbert space
transitions, in a specific limiting procedure as $N \rightarrow \infty$, from a type-I von Neumann algebras to type-III von Neumann algebra \cite{PhysRevD.108.086019, PhysRevD.108.086020}. We will study this phase transition
from a different algebraic perspective linking it to the modular data of nested causal diamonds, using the modular nuclearity condition. 

Another phase transition of interest in holographic theories is the one concerning correlations, which is the mutual information phase transition. This occurs when the mutual information between two disjoint intervals in CFT switches from zero to non-zero. Correspondingly, in the bulk gravitational description, the entanglement wedge transitions from a disconnected geometry to a connected one. This phase transition is controlled by conformally invariant cross-ratios $(z, \bar{z})$ defined by the two boundary subregions. 
Using modular theory, the geometrical data for the vacuum subregions can be related to the thermal partition function, where we can assign a temperature to the mutual information phase transitions $T_{MI} (z, \bar{z})$ in the vacuum state and compare it with $T_{HP}$.

We will also consider the Hagedorn transition \cite{osti_4562936}, which is relevant to stringy physics occurring at the stringy temperature scale $T_s$ setting the
exponential growth of the density of string states. This transition is usually hidden under the Hawking-Page transition, but by considering only perturbative degrees of freedom about the thermal AdS background, this effect becomes visible. Another goal of this paper is to see signatures of the Hagedorn transition in the modular nuclearity partition function. We can then formally access this by taking a strict large N limit, working below $T_{HP}$ or $T_{MI}$, and then continuing the answer above these temperatures to $T_s$. Surprisingly, the Hagedorn transition is set by the distance between the disjoint entanglement wedges in string units. This motivates us to use this as a new probe of entanglement wedges, allowing us to define a new quantity that measures the minimal distance between entanglement wedges in string units. 


The paper is structured as follows. We start by looking at the relevant properties of nested causal diamonds in QFT, specifically nuclearity and split property, which allows one to have statistically independent descriptions of disjoint subregions. Despite the fact that there are quite a few nuclearity conditions formulated in the literature, the relevant one for this paper will be $L_2$ Nuclearity, which we review in Section \ref{sect:2}. In Sections \ref{sect:3}-\ref{sect:4}, we provide our two bounds on mutual information in terms of the thermal partition function. Section \ref{sect:5} focuses on relating the conformal cross-ratio associated with the causal diamonds with the temperature of the partition function, while in Section \ref{sect:6}, we explicitly derive this temperature using AdS/CFT duality following the work of \cite{Jokela_2019}. In Section \ref{sect:7}, we compute and compare the critical temperatures at which the two phase transitions occur. In Section \ref{sect:8}, we compare the mutual information with the logarithm of partition function for the holographic case, namely the AdS vacuum, where the bound on the mutual information is satisfied. In Section \ref{sect:9}, we make a conjecture about the gravitational dual of the inverse temperature and the angular velocity term that goes into the expression for the partition function for general subregions in CFTs. In Section \ref{sect:10}, we make a conjecture about a quantity that relates the Hagedorn transition to the bulk distance between the causal wedges of the subregions for general holographic backgrounds and comment on avenues for future work. We leave the relevant properties of the relative entropy, which we use to prove our bounds, to Appendix \ref{sect:A}. In Appendix \ref{sect:B}, we provide the calculation of the relevant bulk quantities for the example of $AdS_3/CFT_2$.

\section{\texorpdfstring{$L_2$}{L2} Nuclearity and Split Property}
\label{sect:2}

In this section, we review nuclearity conditions and the split property in Quantum Field Theory. The split property states that given algebras associated with two disjoint causal diamonds $\mathcal{A}$ and $\mathcal{B}'$, whose causal completion $\mathcal{B}$ contains $\mathcal{A}$, one then has interpolating type-I factors denoted by $\mathcal{N}$ contained between the algebra associated with $\mathcal{B}'$ and that of $\mathcal{A}$ \cite{Buchholz:1989zz}:\footnote{because $\mathcal{A},\mathcal{B}$ are local quantum field theory von Neumann algebras it is natural to take them to be both in standard form (i.e. there exists a cyclic and separating vector) on the same Hilbert space. We will assume this to be the case throughout the paper.}

\begin{equation}
\mathcal{A} \subset \mathcal{N} \subset \mathcal{B'}
\end{equation}

\noindent
This intermediate type-I factor is infinite-dimensional and there are infinitely many interpolating type-I factors between the two algebras. Unlike the algebras $\mathcal{A}$ and $\mathcal{B}'$, the algebra $\mathcal{N}$ does not correspond to a well-defined space-time region. The existence of $\mathcal{N}$ can also be equivalently formulated as the presence of an isomorphism between the union of type III algebras $\mathcal{A} \vee \mathcal{B}$ and the product of tensor factors $\mathcal{A} \otimes \mathcal{B}$, which is a statement of statistical independence between subregions. The split property itself follows from nuclearity. Although there are numerous formulations of nuclearity in the literature, we will specifically focus on what is known as $L_2$-modular nuclearity, for which the role of modular inclusions becomes manifest. Our starting point will be the \textit{trace-class condition}, which can be formulated in terms of the finiteness of the partition function:

\begin{equation}
Z(\beta,\Omega) = \Tr_{\mathcal{H}} \sqrt{X^{\dag} X}  < \infty
\quad \Longrightarrow \quad X= \Delta_{\mathcal{A}}^{-1/4} \Delta_{\mathcal{B}'}^{1/4} = \sum_i \lambda_i \ket{y_i}  \bra{x_i}
\end{equation}

\noindent 
where we related the modular data of the causal diamonds corresponding to the subregions of QFT vacuum to the partition function of the theory. We defer the discussion of the relationship between geometry and thermodynamic variables (temperature and angular frequency) to Section \ref{sect:5}. Once we have the $L_2$ nuclearity condition, which leads to the trace class condition for the partition function and vice versa, the split property follows \cite{Buchholz:1989zz}, allowing us to define a localization map between a single-copy and double-copy Hilbert spaces and a homomorphism between the algebra of the disjoint unions and tensor products: 

\begin{equation}
U : \mathcal{H} \rightarrow \mathcal{H} \otimes \mathcal{H} \quad , \quad \Phi: \mathcal{A} \vee \mathcal{B} \rightarrow \mathcal{A} \otimes \mathcal{B} 
\end{equation}

\noindent 
It is imperative to have a buffer zone that separates the two algebras for the localization map to exist. This allows us to write the expectation values of the operators belonging to the disjoint union of the algebras evaluated for the vacuum state in the algebras of the tensor product. Although this map is not unique, there is a canonical choice that also fixes the interpolating type-I factor, using the standard representation \cite{Doplicher1984}:

\begin{equation}
\bra{\Omega} a b \ket{\Omega} = \bra{\xi_\Omega} \Phi (a b) \ket{\xi_\Omega} \quad, \quad \xi_{\Omega} \in \mathcal{P}^{\natural} \subset \mathcal{H} \otimes \mathcal{H}
\end{equation}

\noindent
On the tensor product Hilbert Space, the vacuum $\ket{\Omega}$ becomes $\ket{\xi}$, where, following the work of Hollands and Sanders \cite{hollands2018entanglement}, we have the following decomposition into normal linear functionals:

\begin{equation}
\begin{aligned}
\bra{\Omega} a b \ket{\Omega} &= \sum_i \omega_i(ab)
= \sum_i \lambda_i \phi_i (a) \psi_i(b) \\
&= \sum_i \lambda_i \bra{\Omega} a \Delta_{\mathcal{A}}^{1/4} \ket{y_i}  \bra{x_i} \Delta_{\mathcal{B}}^{1/4} b\ket{\Omega}
\end{aligned}
\end{equation}

\noindent
The fact that $\phi_i$ and $\psi_i$ are linear functionals can be seen from the following:

\begin{equation}
\begin{aligned}
\left| \bra{\Omega} a \Delta^{1/4}_\mathcal{A} \ket{y_i} \right| &\leq \sqrt{\bra{\Omega} a^{\dag} \Delta^{1/2}_\mathcal{A} a \ket{\Omega}}  \| \ket{y_i} \| \\
&= \sqrt{\bra{\Omega} a^{\dag} J_\mathcal{A} a^{\dag} J_\mathcal{A} \ket{\Omega}} \| \ket{y_i} \| \\
&\leq \|a\| \| \ket{y_i} \|
\end{aligned}
\end{equation}

\noindent
Here, $\ket{x_i}, \ket{y_i}$ form an orthonormal basis, and $\sum_i \lambda_i < \infty$.  We can use the existence of a unique polar decomposition on the non-commutative $L^1(\mathcal{A}) = \mathcal{A}_\star$ space; see, for example, Araki-Masuda \cite{ArakiMasuda1982}, to write

\begin{equation}
\bra{\Omega} a \Delta_\mathcal{A}^{1/4} \ket{y_i} = \bra{\eta_i} a w_i \ket{\eta_i} \hspace{1cm} \bra{x_i} \Delta_\mathcal{B}^{1/4} b\ket{\Omega} = \bra{\xi_i} b v_i \ket{\xi_i}
\end{equation}

\noindent 
where $w_i \in \mathcal{A}$ and $v_i \in \mathcal{B}$ are unitaries and $\eta_i$ and $\xi_i$ are subnormalized vectors. We then define the following normalized separable linear functional:

\begin{equation}
\hat{\rho} = \sum_i \frac{1}{2} p_i (\omega_{\hat{\eta}_i} \rvert_{\mathcal{A}} \otimes \omega_{\hat{\xi}_i} \rvert_{\mathcal{B}} + \omega_{w_i\hat{\eta}_i} \rvert_{\mathcal{A}} \otimes \omega_{v_i\hat{\xi}_i} \rvert_{\mathcal{B}})
\end{equation}

\noindent 
where $\hat{\eta_i} = \frac{\eta_i}{\|\eta_i\|} , \hat{\xi_i} = \frac{\xi_i}{\|\xi_i\|}$ are normalized vectors. We also have $p_i = \lambda_i \lVert \eta_i \rVert^2 \lVert \xi_i \rVert^2 /Z$ with $Z = \sum_i \lambda_i \lVert \eta_i \rVert^2 \lVert \xi_i \rVert^2 < \infty$. Following \cite{hollands2018entanglement,panebianco2022modular}, we can bound the linear functional corresponding to the vacuum by the separable one. We begin with the following unnormalized linear functional:

\begin{equation}
\rho_i(\cdot) = \frac{1}{2} \sigma_i (\cdot) + \frac{1}{2} \sigma_i (v^{\dag} \otimes w^{\dag} \cdot v \otimes w )
\end{equation}

\noindent where we defined $\sigma_i = \lambda_i \bra{\eta_i} \cdot \ket{\eta_i} \bra{\xi_i} \cdot \ket{\xi_i}$. 
Taking into account the individual terms in the sum of (2.9), $\omega_i = \lambda_i \phi_i \otimes \psi_i$, we can bound the difference with the separable $\rho_i$:

\begin{equation}
\begin{aligned}
0 \leq \frac{1}{2}\sigma_i((1-w^{\dag} \otimes v^{\dag})\cdot(1-w \otimes v)) = \rho_i(\cdot) - \frac{1}{2}(\omega_i(\cdot) + \omega_i^\star(\cdot))
\end{aligned}
\end{equation}
and this implies
that the sum $\omega = \sum_i \omega_i$ (which is itself real and positive) satisfies:
\begin{equation}
 \quad \Rightarrow \quad 0 \leq Z \hat{\rho} - \omega
\end{equation}
\noindent 
Normalizing the separable linear functional $\rho = \sum_i \rho_i$, we can express $\omega$ as follows:

\begin{equation}
\omega = \|\rho\| \hat{\rho} - \|\delta\| \hat{\delta}
\end{equation}

\noindent where we have $\rho(1) = \|\rho\| = \sum_i \lambda_i \| \eta_i \|^2  \| \xi_i \|^2 = Z$ and $\omega(1) =1$. This gives rise to the following inequality:

\begin{equation}
\omega_{\xi} \rvert_{\mathcal{A} \otimes \mathcal{B}} \leq Z \hat{\rho}
\end{equation}






\noindent 
We use the separable state $\hat{\rho}$ and the above inequality to provide bounds on mutual information in the following sections.  


\section{First Estimate}
\label{sect:3}

\noindent
In this section, we provide our first upper bound on the mutual information between vacuum subregions in the QFT. Our bound is similar to the one derived in Theorem 25 of \cite{panebianco2022modular}, where the authors use the partition function constructed from the modular $p-$nuclearity condition to derive their bounds. In contrast, our starting point is the trace-class and $L_2$ nuclearity condition. We begin with the separable state $\hat{\rho}$ from the previous section, which can be written as a linear combination of a linear functional corresponding to the vacuum acting on the tensor product Hilbert Space $\mathcal{A} \otimes \mathcal{B}$ and another linear functional denoted by $\delta$:

\begin{equation}
\hat{\rho} = \frac{1}{Z} \omega_{\xi} \rvert_{\mathcal{A \otimes \mathcal{B}}}  + \delta
\end{equation}

\noindent 
where $\delta \in (\mathcal{A} \otimes \mathcal{B})^{+}_{\star}$ has normalization $\delta(1) = 1- 1/Z$. Using the super-additivity property of relative entropy with respect to its first argument or Donald's identity shown in the appendix \ref{sect:A}, we have the following.

\begin{equation}
\begin{aligned}
I(\mathcal{A} \otimes \mathcal{B})_{\hat{\rho}} &= S(\hat{\rho}| \phi_A \otimes \phi_B) = S(\frac{1}{Z} \omega + \delta | \phi_A \otimes \phi_B) \\
&\geq \frac{1}{Z}S(\omega | \phi_A \otimes \phi_B) + \left( 1- \frac{1}{Z} \right)S( \hat{\delta} | \phi_A \otimes \phi_B) - H \left(\left\{\frac{1}{Z},1-\frac{1}{Z} \right\} \right)
\end{aligned}
\end{equation}

\noindent
By dropping the second term, multiplying both parts by Z and minimizing the relative entropy over states $\phi_A \in \mathcal{A}_{\star}$ and $\phi_B \in \mathcal{B}_{\star}$, we obtain the following upper bound on mutual information.

\begin{equation}
S(\omega| \phi_A \otimes \phi_B) \leq Z I(\mathcal{A} \otimes \mathcal{B})_{\hat{\rho}} + Z H \left(\left\{\frac{1}{Z},1-\frac{1}{Z}\right\} \right)
\end{equation}

\noindent 
We can upper bound the mutual information for the separable state using the following monotonicity relation:

\begin{equation}
I(\mathcal{A} \otimes \mathcal{B})_{\hat{\rho}} \leq I ((\mathcal{B}(\mathcal{H})\otimes \ell_2(\mathbb{N}) ) \otimes (\mathcal{B}(\mathcal{H}) \otimes \ell_2(\mathbb{N})))_R
\end{equation}

\noindent
where we define R to be the following separable state:

\begin{equation}
R= \sum_{i,j} p'_{ij} \left( \omega_{\hat{\xi}'_i} \otimes \ket{i}\bra{i} \otimes \omega_{\hat{\eta}'_j} \otimes \ket{j} \bra{j} \right) = \sum_i p_i'\left(\omega_{\hat{\xi}'_i} \otimes \ket{i}\bra{i} \otimes \omega_{\hat{\eta}'_i} \otimes \ket{i} \bra{i} \right)
\end{equation}

\noindent 
where we converted the double sum into a single sum with twice as many terms with $p'_{1+2i}= p'_{2i} = \frac{p_i}{2}$, using the fact that we have a classically correlated state composed of orthonormal states. Using the definition of mutual information in terms of the entanglement entropies evaluated for the respective algebras, we have: $I_R = S\left(R\big|_{\mathcal{A}}\right) + S\left(R\big|_{\mathcal{B}}\right) - S(R \big|_{\mathcal{A \otimes \mathcal{B}}})$. For a classical mixture, we see that all three terms contribute the same Shannon piece, where the mutual information for the extended state becomes:

\begin{equation}
I ((\mathcal{B}(\mathcal{H})\otimes \ell_2(\mathbb{N}) ) \otimes (\mathcal{B}(\mathcal{H}) \otimes \ell_2(\mathbb{N})))_R = \sum_i - p_i \log \frac{p_i}{2}
\end{equation}

\noindent 
Hence, we have the following bound on the mutual information for the separable state $\hat{\rho}$:

\begin{equation}
I(\mathcal{A} \otimes \mathcal{B})_{\hat{\rho}} \leq \sum_i - p_i \log \frac{p_i}{2}
\end{equation}

\noindent
Using this monotonicity bound, we have the estimate:

\begin{equation}
\begin{aligned}
I(\mathcal{A} \otimes \mathcal{B})_{\omega_{\xi}} &\leq Z \left(-\sum_i p_i \log \frac{p_i}{2} \right) + Z H\left(\left\{\frac{1}{Z},1-\frac{1}{Z}\right\}\right)
\end{aligned}
\end{equation}

\noindent
Similarly to the result derived in \cite{panebianco2022modular}, this bound contains terms that are linear in the partition function itself. In the next section, we will describe a second, stronger bound on mutual information.

\section{Second Estimate}
\label{sect:4}

This stronger estimate makes use of the max-relative entropy. Using the monotonicity property of relative entropy with respect to Renyi parameter, we start with the fact that the max-mutual information is greater than the mutual information for the vacuum state for subalgebras $\mathcal{A}$ and $\mathcal{B}$, using the variational expression for max-mutual information outlined in \cite{Ciganovic_2014}:

\begin{equation}
\begin{aligned}
I(\mathcal{A}\otimes \mathcal{B})_{\omega_{\xi}} &\leq I_{max}(\mathcal{A}\otimes \mathcal{B})_{\omega_{\xi}} \\
I_{max}(\mathcal{A}\otimes \mathcal{B})_{\omega_{\xi}} &= \inf_{\phi_A,\phi_B} \inf_{\lambda} \left\{ \log \lambda  : \omega_{\xi}|_{\mathcal{A}\otimes \mathcal{B}} < \lambda \phi_A \otimes \phi_B     \right\}
\end{aligned}
\end{equation}

\noindent
where $\phi_A$ and $\phi_B$ are taken as normalized linear functionals. We upper bound $I_{max}^{\omega_\xi}$, by using the max-mutual information for a known separable linear functional $\rho$, for which we have the following:

\begin{equation}
I_{max}(\mathcal{A} \otimes \mathcal{B})_{\hat{\rho}} = \inf_{\phi_A,\phi_B} \inf_{\lambda} \left\{ \log \lambda  : \hat{\rho}|_{\mathcal{A}\otimes \mathcal{B}} < \lambda \phi_A \otimes \phi_B     \right\}
\end{equation}

\noindent
Using the inequality between the linear functionals derived previously, $\omega_{\xi}|_{\mathcal{A}\otimes \mathcal{B}} < Z \hat{\rho}$, we have the following relation:

\begin{equation}
\begin{aligned}
I_{max}(\mathcal{A} \otimes \mathcal{B})_{\hat{\rho}} + \log Z  &= \inf_{\phi_A,\phi_B} \inf_{\lambda'} \left\{ \log (Z \lambda')  : Z \hat{\rho}  < Z \lambda' \phi_A \otimes \phi_B     \right\} \\
&\geq I_{max}(\mathcal{A} \otimes \mathcal{B})_{\omega_{\xi}}
\end{aligned}
\end{equation}

\noindent
The above inequality follows from the fact that the infimum is realized over a smaller set of separable linear functionals and $\lambda$ compared to the one on the right side of the inequality. This gives the following:

\begin{equation}
I(\mathcal{A} \otimes \mathcal{B})_{\omega_{\xi}} \leq I_{max}(\mathcal{A} \otimes \mathcal{B})_{\omega_{\xi}}  
\leq I_{max} ((\mathcal{B}(\mathcal{H})\otimes \ell_2(\mathbb{N}) ) \otimes (\mathcal{B}(\mathcal{H}) \otimes \ell_2(\mathbb{N})))_R + \log Z
\end{equation}

\noindent
where we also used the property of monotonicity for max-relative entropy with respect to restrictions/extensions of the algebras, and the orthogonal separable state $R$ from before. Since the mutual information for the orthogonal separable state is given by the Shannon entropy of a classical probability distribution with only diagonal terms contributing, to obtain the max-relative entropy, one needs the largest eigenvalue:

\begin{equation}
I_{max} ((\mathcal{B}(\mathcal{H})\otimes \ell_2(\mathbb{N}) ) \otimes (\mathcal{B}(\mathcal{H}) \otimes \ell_2(\mathbb{N})))_R \leq \min_i \left\{-\log \frac{p_i}{2} \right\}
\end{equation}

\noindent
From the first estimate, the probability densities are defined as $p_i = \frac{\lambda_i \lVert \xi_i \rVert^2 \lVert \eta_i \rVert^2}{Z}$, where $\lambda_i = e^{-\beta(E_i - E_0)}$ and the maximum probability can be attained by the eigenvalue corresponding to the vacuum: $p_{max} = \frac{\lVert \xi_0 \rVert^2 \lVert \eta_0 \rVert^2}{Z}$. This results in the following estimate:

\begin{equation}
I_{max}(\mathcal{A} \otimes \mathcal{B})_{\omega_{\xi}} \leq \log(2Z^2) - 2 \log \Vert \xi_0 \rVert \Vert \eta_0 \rVert 
\end{equation}

\noindent
Since the corresponding vector is vacuum, in this case we have:

\begin{equation}
\ket{x_0}, \ket{y_0} = \ket{\Omega} \quad  \ket{\eta_0}, \ket{\xi_0} = \ket{\Omega} \quad w_0, v_0 = 1
\end{equation}

\begin{equation}
I_{max}(\mathcal{A} \otimes \mathcal{B})_{\omega_{\xi}} \leq \log 2 + 2 \log Z \leq \log2 + 2\log Z(\beta)
\end{equation}

\noindent
where we used the fact that the norms of the linear functionals $\Vert \xi_0 \rVert$ and $\Vert \eta_0 \rVert$ are equal to 1, corresponding to the vacuum state, and the rest of the states in the sum are subnormalized. In the latter inequality, we used the fact that the thermal partition function $Z(\beta) = \sum_i e^{-\beta (E_i - E_0)} \geq Z = \sum_i e^{-\beta (E_i - E_0)} \Vert \xi_i \rVert^2 \Vert \eta_i \rVert^2 $. Expressing the thermal partition function in terms of the free energy and normalizing with respect to the vacuum contribution, the Casimir energy, we end up with the following final result: 

\begin{equation}
I(\mathcal{A}\otimes \mathcal{B})_{\omega_{\xi}} \leq \log2 + 2\log Z(\beta) = \log2 -2 \beta (F - F_0)
\end{equation}

\noindent
The second estimate is stronger than the first, as it just involves the logarithm of the partition function as opposed to the partition function itself. We will now move to the construction of the partition function from the modular data of the causal diamonds before showing that the bound we derived is satisfied for the specific example of spherical subregions in holographic CFTs.

\section{Diamonds}
\label{sect:5}

\noindent 
We work in the embedding space, which is the $d+1$ dimensional, codimension one, submanifold: 

\begin{equation}
-(P^0)^2 - (P^1)^2 + (P^2)^2 + \hdots (P^{d+1})^2 = 0
\end{equation}

\noindent
of the projective plane with metric: 
\begin{equation}
ds^2 = -(dP_0)^2 -(dP_1)^2 + (dP_2)^2 + \hdots (dP_{n+1})^2 \equiv \eta_{MN} dP^{M} dP^{N}
\end{equation}

\noindent
and the identification that $P^M \cong \lambda P^M$. We are mostly interested in working on static cylinder, which is parameterized with:

\begin{equation}
P^0 = \cos t  \hspace{1cm} P^1 = \sin t \hspace{1cm} P^{1+i} = n^i
\end{equation}

\noindent
where $\sum_{i=1}^d (n^i)^2 = 1$ gives the unit $d-1$ sphere. The $\mathcal{B}'$ region has the modular group determined by: 

\begin{equation}
\begin{pmatrix}
P^1 \\
P^{d+1}
\end{pmatrix}' =  
\begin{pmatrix}
\cosh 2\pi s &\sinh 2\pi s \\
\sinh 2\pi s &\cosh 2\pi s \\
\end{pmatrix}
\begin{pmatrix}
P^1 \\
P^{d+1}
\end{pmatrix}
\end{equation}

\noindent 
which holds fixed the rays:

\begin{equation}
P^M_{\pm} (\mathcal{B}') = \lambda (0,\pm 1,0,\hdots,1) 
\end{equation}

\noindent 
On the cylinder this acts as:

\begin{equation}
\begin{pmatrix}
\sin t \\
n^d
\end{pmatrix}' = \lambda
\begin{pmatrix}
\cosh 2\pi s &\sinh 2\pi s \\
\sinh 2\pi s &\cosh 2\pi s \\
\end{pmatrix}
\begin{pmatrix}
\sin t \\
n^d
\end{pmatrix} \hspace{1cm} \cos t' = \lambda \cos t \hspace{1cm} (n^i)' = \lambda n^i \lvert_{i\neq d}
\end{equation}

\noindent 
where, we must pick $\lambda$ to preserve the gauge choice. Solving for $\lambda$, we find that

\begin{equation}
\lambda^{-2} = (\cosh (2\pi s)\sin t + \sinh (2\pi s)n^d )^2 + \cos^2 (t)
\end{equation}

\noindent 
and the locations of the fixed points are:

\begin{equation}
t= \pm \pi/2, \hspace{1cm} n^d =1, \hspace{1cm} n^{i\neq d} = 0
\end{equation}

\noindent 
Now for the smaller subdiamond $\mathcal{A}$, we would like to act on $\mathcal{B}'$ with what would be a dilation about the point $n^d=1, t=0$ in the corresponding Poincare patch. We also would like to apply a conformal boost associated with a causal diamond that is rotated in the $(d,d-1)$ plane for the sphere. These two commute. This will give a general enough diamond to cover all possibilities. We apply the following transformation:

\begin{equation}
\begin{pmatrix}
P^0 \\
P^1 \\
P^d \\
P^{d+1}
\end{pmatrix}' = 
\begin{pmatrix}
\cosh p_1 & 0 & 0 & \sinh p_1 \\
0 & \cosh p_2 & \sinh p_2 & 0 \\
0 & \sinh p_2 & \cosh p_2 & 0 \\
\sinh p_1 & 0 & 0 & \cosh p_1  
\end{pmatrix}
\begin{pmatrix}
P^0 \\
P^1 \\
P^d \\
P^{d+1}
\end{pmatrix}
\end{equation}

\noindent 
to $\mathcal{B}'$. The new diamond has future and past fixed points now at the rays:

\begin{equation}
P^{M}_{\pm} (\mathcal{A}) = \lambda (\sinh p_1, \pm \cosh p_2, 0, \hdots, \pm \sinh p_2, \cosh p_1) 
\end{equation}

\noindent 
if we pick $\lambda=1/\sqrt{\sinh^2 p_1 + \cosh^2 p_2}$, then this is the cylinder frame where we could work out the end points of the diamond. We can work out the cross-ratios using:

\begin{equation}
x^2_{12} = - 2 P_1 \cdot P_2
\end{equation}

\noindent 
We choose the distances to be $x_{1,2} = P_{\pm}(\mathcal{B})$ and $x_{3,4} = P_{\mp}(\mathcal{A})$. Then we find (dropping the overall $\lambda$ terms in the cross ratio):

\begin{equation}
\begin{aligned}
P_1 \cdot P_2 = P_3 \cdot P_4 &= 2  \hspace{1cm} P_1 \cdot P_3 = P_2 \cdot P_4 = \cosh p_1 + \cosh p_2 \\
P_1 \cdot P_4 &= P_2 \cdot P_3 = - \cosh p_2 + \cosh p_1
\end{aligned}
\end{equation}

\noindent 
so that we have the following:

\begin{equation}
u= \frac{4}{(\cosh p_1 + \cosh p_2)^2} \hspace{1cm}   v= \frac{(\cosh p_1 - \cosh p_2)^2}{(\cosh p_1 + \cosh p_2)^2}
\end{equation}

\begin{equation}
z= \frac{1}{\cosh^2((p_1+p_2)/2} \hspace{1cm}   \bar{z} = \frac{1}{\cosh^2((p_1-p_2)/2)}
\end{equation}

\noindent
Now, we can work out the expression for the conformal generators following \cite{hollands2018entanglement,2007CMaPh.270..267B}. These are given in terms of the modular operators associated with the causal diamonds:

\begin{equation}
\Delta^{is}_{\mathcal{A}} \Delta^{-is}_{\mathcal{B}'} = u^{\dag} \Delta^{is}_{\mathcal{B}'} (u) \Delta^{-is}_{\mathcal{B}'}
\end{equation}

\noindent 
We note that $\Delta^{is}_{\mathcal{B}'} (u) \Delta^{-is}_{\mathcal{B}'}$ is then just the conformal boost of $u= u(p_1,p_2)$, which is the unitary representation of the transformation that takes one diamond to the other. We can write this as follows:

\begin{equation}
\Delta^{is}_{\mathcal{A}} \Delta^{-is}_{\mathcal{B}'} = u^{\dag} \exp(ip_1 \Delta^{is}_{\mathcal{B}'} Q_1 \Delta^{-is}_{\mathcal{B}'} + ip_2 \Delta^{is}_{\mathcal{B}'} Q_2 \Delta^{-is}_{\mathcal{B}'} )
\end{equation}

\noindent 
where $Q_{1,2}$ are the generators of the two transformations described above. The generators correspond to 

\begin{equation}
Q_1 \leftrightarrow i (P^0 \partial_{P^{d+1}}+ P^{d+1} \partial_{P^0})  \hspace{1cm} Q_2 \leftrightarrow i (P^1 \partial_{P^d} + P^d \partial_{P^1})
\end{equation}

\noindent 
Applying the boost gives:

\begin{equation}
\begin{aligned}
R_{\mathcal{B}'}(2\pi s)^{-1}Q_1R_{\mathcal{B}'}(2\pi s) \rvert_{s=i/4} &\leftrightarrow i (P^0 \partial_{P^{d+1}}- P^{d+1} \partial_{P^0}) \\
R_{\mathcal{B}'}(2\pi s)^{-1} Q_2 R_{\mathcal{B}'}(2\pi s) \rvert_{s=i/4} & \leftrightarrow i (P^1 \partial_{P^d} - P^d \partial_{P^1})
\end{aligned}
\end{equation}

\noindent
so that the final expression is:

\begin{equation}
\Delta^{-1/4}_{\mathcal{A}} \Delta^{1/4}_{\mathcal{B'}} = u^{\dag} \exp(-p_1 H -p_2 J )
\end{equation}

\noindent 
where $J$ is the rotation generator of the sphere in the $(d, d-1)$ plane of the $n^{i}$ coordinates. Therefore, $p_1= \beta$ and $p_2= \beta \Omega$, where $\Omega$ is the angular velocity of the fluid in this direction. We can relate the inverse temperature and the angular velocity to the conformal cross ratios. We have

\begin{equation}
z= \frac{1}{\cosh^2(\beta(1+\Omega)/2)}  \hspace{1cm} \bar{z} = \frac{1}{\cosh^2(\beta(1-\Omega)/2)}
\end{equation}

\noindent 
Alternatively, one can write the following:

\begin{equation}
\begin{aligned}
\beta &= \cosh^{-1}(z^{-1/2}) + \cosh^{-1}(\bar{z}^{-1/2}) \\
\beta \Omega &= \cosh^{-1}(z^{-1/2}) - \cosh^{-1}(\bar{z}^{-1/2})
\end{aligned}
\end{equation}

\noindent 
In $2d$, we can relate the cross ratios to the left and right temperatures:

\begin{equation}
z= \frac{1}{\cosh^2(\beta_{L}/2)} \hspace{1cm} \bar{z} = \frac{1}{\cosh^2(\beta_{R}/2)}
\end{equation}

\noindent 
Having related the conformal cross ratios to the temperature and the partition function, we now proceed to the specific example of the d-dimensional holographic CFTs and compute the inverse temperature for annular regions. 

\section{Spherical Subregions in Holographic CFTs}
\label{sect:6}

\noindent
This section mainly follows the results of \cite{Jokela_2019} and also the earlier results on mutual information for holographic theories in higher dimensions studied in \cite{Hirata_2007} and later in \cite{Fonda_2015}. We consider a spherically symmetric RT surface parameterized by $z(\rho)$. The pure AdS metric in spherical coordinates is given by

\begin{equation}
\frac{L^2_{AdS}}{z^2} \left(-dt^2 + dz^2 + d\rho^2 + \rho^2 d \Omega^2_{d-2} \right)
\end{equation}

\noindent 
We consider equal-time solutions, since all time-dependent spherically symmetric solutions can be obtained via a conformal transformation of the equal-time solution. We start with the following area functional: 

\begin{equation}
A= \int \frac{d\rho \rho^{d-2}}{z^{d-1}} \sqrt{1+z'(\rho)^2}
\end{equation}

\noindent
where we set the prefactor $L^{d-1}_{AdS} Vol(S^{d-2})$ to 1 for now. If we redefine $\zeta=z/\rho$ and $\tau= t/\rho$ and use $\rho=e^{u}$, then there is now a time translation invariance $u \rightarrow u+c$, resulting in the conserved energy $Q$. The area functional in new variables and the equation for the conserved quantity are given as follows:

\begin{equation}
A = \int du \zeta^{1-d} \sqrt{1+(\zeta+\zeta')^2}
\end{equation}

\begin{equation}
-\frac{1+\zeta(\zeta+\zeta')}{\zeta^{(d-1)}\sqrt{1+(\zeta+\zeta')^2}} = Q
\end{equation}

\noindent
Following \cite{Jokela_2019}, one can solve for $\zeta$ to find the boundary values of the radii $\rho_{\pm}= R_{\pm}$ of the surface corresponding to $z \rightarrow 0$. The two radii are related by the following integral expression:

\begin{equation}
-\log \frac{R_{-}}{R_{+}} = 2 Q \int_0^{\zeta_*}\frac{d\zeta \zeta^{-(d-1)}}{(1+\zeta^2)\sqrt{1+\zeta^2-Q^2\zeta^{2(d-1)}}}
\end{equation}

\noindent 
$\zeta_*$ is the first positive root of the expression in the denominator of the integral. Knowing the boundaries of the equal-time surfaces, we can determine the tips of the causal diamonds. We set $R_{+}=1$ for the radius of the outer surface for simplicity. Using the tips of the causal diamonds, we can compute the conformal cross ratios. In Lorentzian time both $z$ and $\bar{z}$ should be real (i.e. not complex conjugates of each other.) The cross ratios are defined such that $z, \bar{z} \rightarrow 0$ as $R_{-} \rightarrow 0$. That is $x_{1,2}= (\pm R_{-},0)$ and $x_{3,4}= (\pm 1, 0)$. This gives:

\begin{figure}[t]
  \centering
  \includegraphics[width=0.6\linewidth]{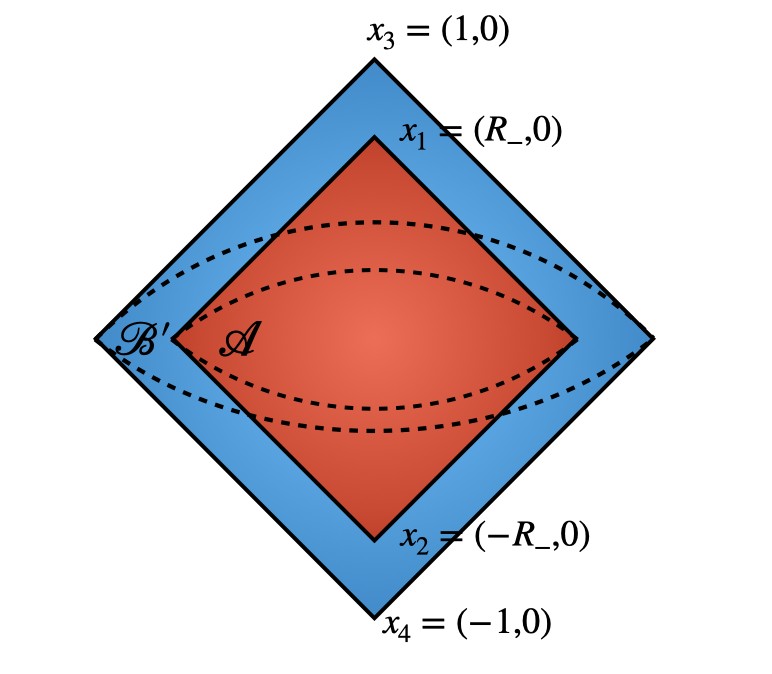}
  \caption{Nested causal diamonds. Invariant cross-ratios are determined from the tips of the lightcones}
  \label{fig:fig2b}
\end{figure}

\begin{equation}
u = \frac{x^2_{12}x^2_{34}}{x^2_{14} x^2_{23}} \hspace{0.5cm}  v= \frac{x^2_{13}x^2_{24}}{x^2_{14}x^2_{23}} \hspace{0.5cm} u = z \bar{z} \hspace{0.5cm} v= (1-z) (1-\bar{z})
\end{equation}

\begin{equation}
u= \frac{16 R^2_{-}}{(1+R_{-})^4} \hspace{1 cm} v= \left(\frac{1-R_{-}}{1+R_{-}}\right)^4
\end{equation}

\noindent
In terms of the integral expression, we have the following final answer for the conformal cross-ratios:

\begin{equation}
z= \bar{z} = \frac{1}{\cosh^2(\beta/2)} = \frac{1}{\cosh^2 \left( Q \int_0^{\zeta_*} \frac{d \zeta \zeta^{(d-1)}}{(1+\zeta^2)\sqrt{1+\zeta^2-Q^2 \zeta^{2(d-1)}}} \right)}
\end{equation}

\noindent
This allows us to extract the inverse temperature corresponding to nested causal diamonds:

\begin{equation}
\beta = 2Q \int_0^{\zeta_*} \frac{d \zeta \zeta^{(d-1)}}{(1+\zeta^2)\sqrt{1+\zeta^2-Q^2 \zeta^{2(d-1)}}}
\end{equation}

\noindent
Having found the expression for the inverse temperature, in the next section, we will look at the critical temperature at which the mutual information transition occurs. 

\section{Critical Temperatures}
\label{sect:7}

\noindent
In this section, we compare the critical temperature at which the Hawking-Page transition occurs $T_{HP}$ to the temperature associated with the mutual information phase transition $T_{MI}$. We want to show that $T_{MI} > T_{HP}$ for holographic CFTs. The inverse temperature corresponding to the mutual information transition can be extracted from the conformal cross ratios of the causal diamonds:

\begin{equation}
\beta_{MI} = 2 Q_{MI} \int^{\zeta_{*}}_0 \frac{d\zeta \zeta^{d-1}}{(1+\zeta^2)\sqrt{1+\zeta^2-Q_{MI}^2\zeta^{2(d-1)}}}
\end{equation}

\noindent 
where $Q_{MI}$ denotes the critical value of the conserved quantity at which mutual information transition occurs, that is, when the area corresponding to the two disconnected spheres and the one corresponding to the connected annulus are equal. Using the solution for $\zeta$, we can find an explicit expression for the area functional \cite{Jokela_2019}:

\begin{equation}
\begin{aligned}
A = \int \frac{d\zeta \zeta^{-(d-1)}}{\sqrt{1+\zeta^2-Q^2\zeta^{2(d-1)}}}
\end{aligned}
\end{equation}

\noindent
The mutual information between the two subregions can be found by subtracting the areas corresponding to connected and disconnected solutions using the RT prescription \cite{Ryu_2006, Hirata_2007}, from which we can extract $Q_{MI}$:

\begin{equation}
\begin{aligned}
I(\mathcal{A}: \mathcal{B}) = (Constant) \hspace{0.1 cm} (\Delta A_{disconnected} - \Delta A_{connected}) &=0 \\
\int^{\zeta_*(Q_{MI})}_0 d \zeta \zeta^{-d+1} \left(\frac{1}{\sqrt{1+\zeta^2-Q^2_{MI}\zeta^{2(d-1)}}}-\frac{1}{\sqrt{1+\zeta^2}}\right) &- \int^{\infty}_{\zeta_*(Q_{MI})} \frac{d\zeta \zeta^{-d+1}}{\sqrt{1+\zeta^2}}= 0
\end{aligned}
\end{equation}

\begin{figure}[h]
  \centering
  \includegraphics[width=0.8\linewidth]{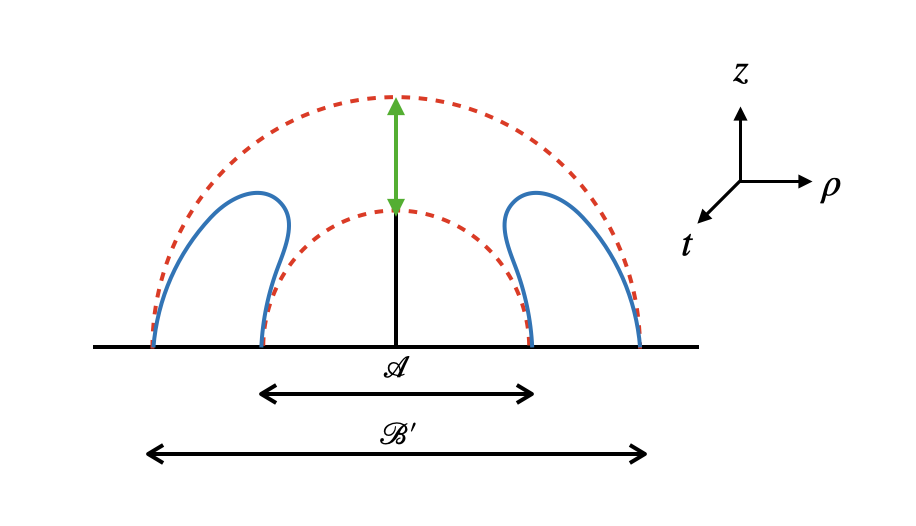}
  \label{fig:fig2}
  \caption{Figure adapted from \cite{Jokela_2019}.  Entanglement wedges corresponding to the two spherical boundary regions (equal-time solutions) associated with diamonds $\mathcal{A}$ and $\mathcal{B}'$. We have the two disconnected spheres solution for $I(\mathcal{A}:\mathcal{B}) =0$ and the deformed annulus solution for $I(\mathcal{A}:\mathcal{B}) >0$. The green arrow represents the minimum distance between the entanglement wedges of $\mathcal{A}$ and $\mathcal{B}$, which also coincides with the inverse temperature $\beta$ for the thermal  partition function. The entanglement wedge cross-section, on the other hand, is the minimum area between the blue surfaces.}
\end{figure}
\noindent

\noindent 
One can show that $\beta_{MI}$ is strictly less than $\beta_{HP}$ by bounding the integral by another, where the conserved quantity $Q_{MI}$ can be scaled away through a change of variables. Using $\frac{1}{1+\zeta^2} < \frac{2}{\zeta}$ for $\zeta >0$ and $\zeta_* \geq Q^{-1/(d-1)}$, we have the following inequalities:

\begin{equation}
\begin{aligned}
\beta_{MI} = 2 Q_{MI} \int^{\zeta_{*}}_0 \frac{d\zeta \zeta^{d-1}}{(1+\zeta^2)\sqrt{1+\zeta^2-Q_{MI}^2\zeta^{2(d-1)}}} &< 4 Q_{MI} \int^{Q_{MI}^{-1/(d-1)}}_0 \frac{d\zeta \zeta^{d-2}}{\sqrt{1-Q_{MI}^2\zeta^{2(d-1)}}} \\
&= 4  \int^{1}_0 \frac{dy y^{d-2}}{\sqrt{1-y^{2(d-1)}}} = \frac{2\pi}{d-1} = \beta_{HP}
\end{aligned}
\end{equation}

\noindent
This analytically proves that the temperature associated with the entanglement wedge phase transition is higher than that of Hawking-Page. For 2 dimensional CFTs, we have $\beta_L \neq \beta_{R}$, and each transition temperature can be described by a one-dimensional curve. The phase diagram of the Hawking-Page transition was studied in \cite{Hartman_2014}. In Figure 3,  we compare it with that of mutual information:

\begin{equation}
\begin{aligned}
\text{Hawking Page:} \quad & \beta_{L} \beta_{R} = 4\pi^2 \\
\text{Mutual Information:} \quad & \frac{1}{\cosh^2\left(\frac{\beta_L}{2}\right)} + \frac{1}{\cosh^2\left(\frac{\beta_R}{2}\right)} = 1
\end{aligned}
\end{equation}

\begin{figure}[htbp!] \hspace{-0.7 cm}
	\begin{subfigure}[b]{0.549\linewidth} 
		\centering
		\includegraphics[width=\textwidth]{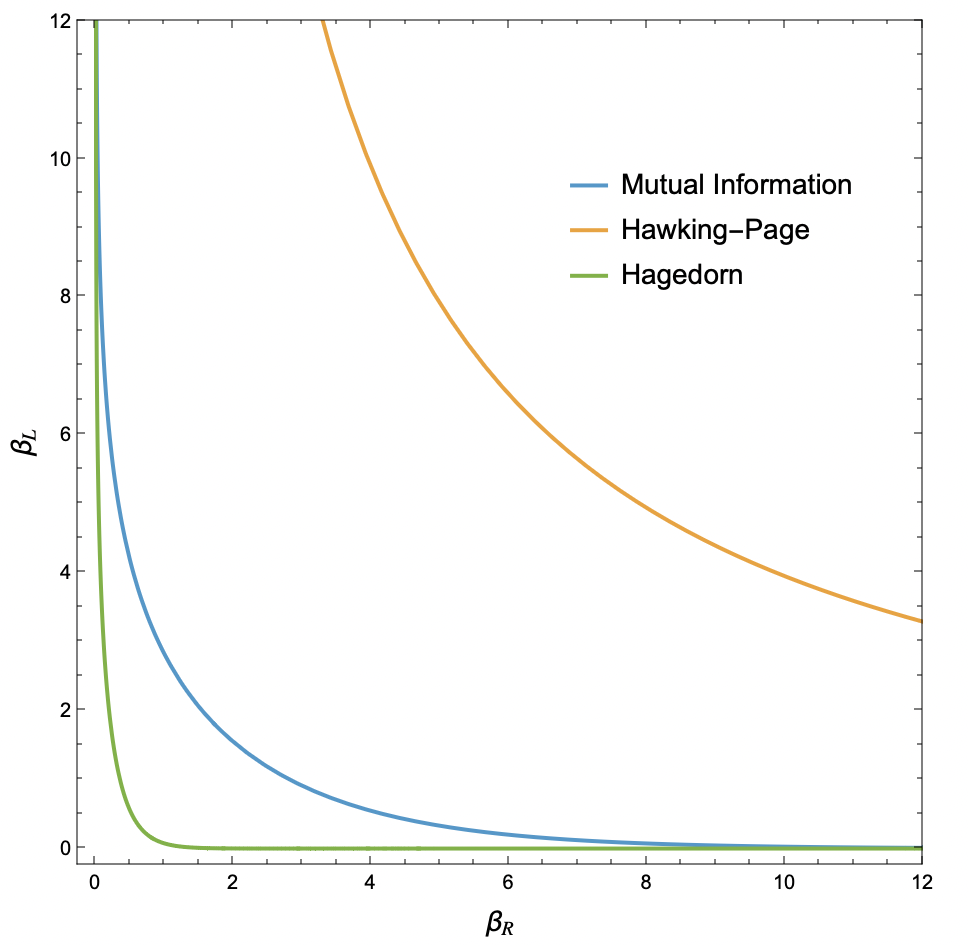} \vspace{-0.42 cm}
	\end{subfigure}\hfill \hspace{-0.25 cm}
	\begin{subfigure}[b]{0.543\linewidth} 
		\centering
		\includegraphics[width= 1.0\textwidth]{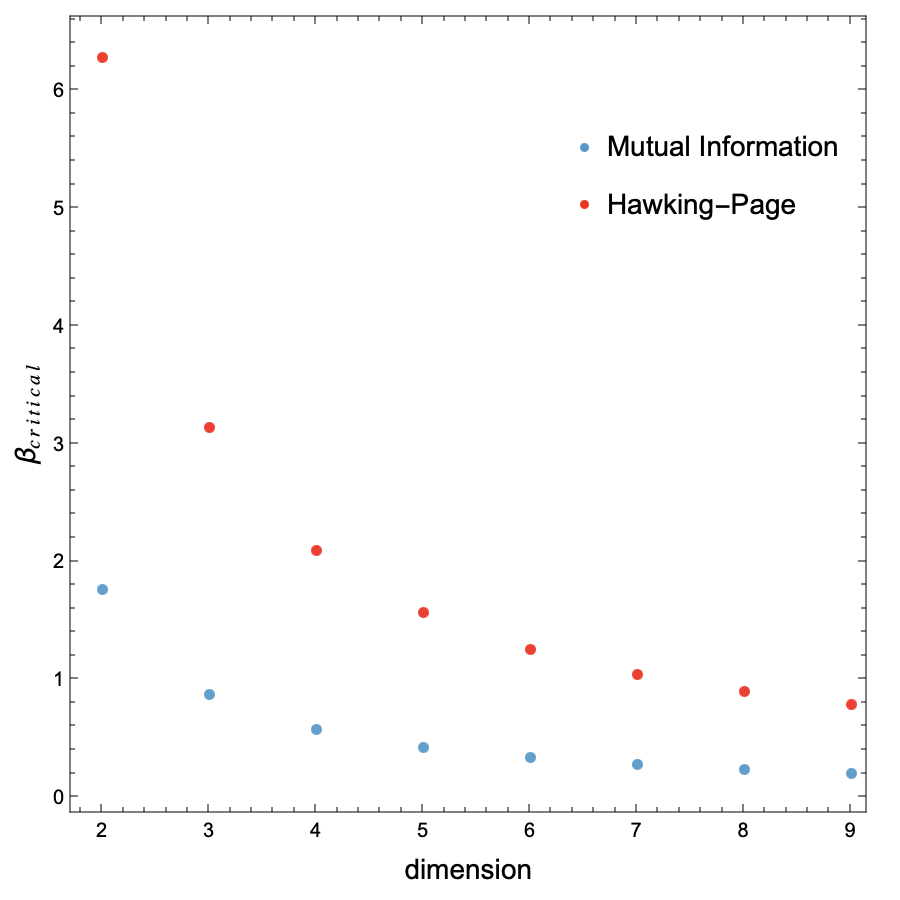}
	\end{subfigure}
	\caption{Right figure: Hawking-Page vs mutual information transition in terms of inverse temperatures as a function of dimension of the CFT, starting at $d=1+1$. Due to the spherical symmetry of the solutions $(d>2)$ we work with, we have $\beta_L = \beta_R$. Left figure: for $d=2$, $\beta_L$ and $\beta_R$ are different, the green curve is a cartoon for Hagedorn transition, which happens at the highest temperature}
	\label{aa}
\end{figure}

\section{High Temperature Dependence of the Partition Function and the Mutual Information}
\label{sect:8}

\subsection{Mutual Information}

In this section, we compare the high-temperature behaviors of the thermal partition function and the mutual information of spherically symmetric subregions in $d$ dimensional holographic CFTs. Using the expression for $\beta$ derived previously, mutual information can be easily related to the inverse temperature:

\begin{equation}
\begin{aligned}
I(\mathcal{A}:\mathcal{B}) &= \frac{Vol(S^{d-2})L^{d-1}_{AdS}}{2G^{(d+1)}_{N}}\left(-\frac{\beta Q}{2} + \int_0^{\zeta_*} d\zeta \zeta^{-d+1}\left( \frac{1}{\sqrt{1+\zeta^2}} - \frac{\sqrt{1+\zeta^2-Q^2\zeta^{2(d-1)}}}{1+\zeta^2} \right) \right. \\
& \left. + \int_{\zeta_*}^{\infty} \frac{d\zeta \zeta^{-d+1}}{\sqrt{1+\zeta^2}} \right)
\end{aligned}
\end{equation}

\noindent 
The inverse temperature $\beta$ and the conserved quantity $Q$ can be related by the following expression when we are in the connected phase, $I(\mathcal{A}:\mathcal{B}) \geq 0$:

\begin{equation}
\frac{\partial I(\mathcal{A}:\mathcal{B})}{\partial \beta} = - \textit{C} Q(\beta)
\end{equation}

\noindent 
where $C$ is the positive constant associated with the prefactors of mutual information. The derivative of mutual information with respect to the inverse temperature yields the conserved quantity $Q$ associated with the extremal surface. This suggests that mutual information itself can be interpreted as an action term $\log Z(\beta)$. Although the CFT dual of the quantity $Q$ is unclear to us, it seems to give a condition on the positivity of conditional mutual information that depends on the minimal distance between the disconnected surfaces. In the limit $Q \rightarrow \infty$ or equivalently $\beta \rightarrow 0$, the smallest positive root $\zeta_*$ approaches $ Q^{-1/(d-1)} \rightarrow 0$. This allows us to omit the $\zeta^2$ from the integrals, resulting in the following large $Q$ dependence of the inverse temperature and the mutual information:

\begin{equation}
\begin{aligned}
\beta  &= \frac{2\sqrt{\pi}}{d}\frac{\Gamma(\frac{3d-2}{2d-2})}{\Gamma(\frac{2d-1}{2d-2})} Q^{-\frac{1}{d-1}} + \ldots\\
I(\mathcal{A}:\mathcal{B}) &=  \frac{Vol(S^{d-2})L^{d-1}_{AdS}}{4G^{(d+1)}_{N}} \left(\frac{\sqrt{\pi}}{d-2}\frac{\Gamma(\frac{d}{2d-2})}{\Gamma(\frac{2d-1}{2d-2})}-\frac{2\sqrt{\pi}}{d}\frac{\Gamma(\frac{3d-2}{2d-2})}{\Gamma(\frac{2d-1}{2d-2})} \right) Q^{\frac{d-2}{d-1}} + \ldots
\end{aligned}
\end{equation}

\noindent 
In terms of both the inverse temperature and the conserved charge, this is given by the following expression:

\begin{equation}
I(\mathcal{A:\mathcal{B}}) = \left(\frac{Vol(S^{d-2})L^{d-1}_{AdS}}{4G^{(d+1)}_{N}}\right) \frac{\beta Q}{d-2} + \ldots
\end{equation}

\noindent
Rewriting the mutual information in terms of the diamond temperature, in the limit $T \rightarrow \infty$, we have:

\begin{equation}
I(\mathcal{A}: \mathcal{B}) = \left(\frac{L^{d-1}_{AdS}}{4G_N^{(d+1)}}\right) \left(\frac{2^d}{d-2} \left(\frac{\pi}{d}\right)^{d-1}\frac{\Gamma(\frac{2d-1}{2d-2})^{1-d}\Gamma(\frac{3d-2}{2d-2})^{d-1}}{\Gamma(\frac{d-1}{2})}\right) T^{d-2} + \ldots
\end{equation}

\noindent
for $d=2$, the temperature dependence is given by the following (valid for all temperatures):

\begin{equation}
I (\mathcal{A}:\mathcal{B}) \big|_{d=2} = \frac{L_{AdS}}{G^{(2)}_N} \log\left[\csch\left(\frac{1}{2T}\right)\right]
\end{equation}

\noindent 
Meanwhile, following \cite{Jokela_2019}, we can also find an expression for the entanglement wedge cross section in the small separation (high temperature limit) and compare it to the mutual information for any CFT dimension $d \geq 2$. This is given by the following expression:

\begin{equation}
EWCS = \left(\frac{Vol(S^{d-2})L^{d-1}_{AdS}}{4G^{(d+1)}_{N}}\right) \int^{\infty}_{\zeta^*} \frac{ \zeta^{1-d} d\zeta }{\sqrt{1+\zeta^2}} = \left(\frac{Vol(S^{d-2})L^{d-1}_{AdS}}{4G^{(d+1)}_{N}}\right) \left(\frac{Q^{\frac{d-2}{d-1}}}{d-2}\right) + \ldots
\end{equation}

\noindent We can also express the ratio of the two quantities in this limit, which approaches a constant:
\begin{equation}
\frac{EWCS}{I(\mathcal{A}:\mathcal{B})} \rightarrow \frac{d}{2\sqrt{\pi}} \frac{\Gamma(\frac{2d-1}{2d-2})}{\Gamma(\frac{3d-2}{2d-2})}
\end{equation}

\subsection{Partition Function}
Having obtained the high-temperature behavior of the mutual information, we now move to the calculation of the thermal partition function for the holographic CFTs. The expression for the partition function is well known in the literature and can be determined from the dual AdS metric of the black hole geometry. Depending on temperature, one has two different solutions: the thermal AdS phase and the black hole phase separated by the Hawking-Page temperature \cite{Hawking1983ThermodynamicsOB}. The inverse temperature for the black hole metric is determined by eliminating the conical singularity and is given in terms of the radius $r_+$ of the outer horizon of the black hole:

\begin{equation}
\beta = \frac{4 \pi r_+ L^2_{AdS}}{dr^2_{+}+ (d-2)L^2_{AdS}}
\end{equation}

\noindent
Since the mutual information transition is expected to happen at a diamond temperature higher than that of the Hawking-Page transition, we are in the black hole phase when the mutual information is non-zero and one needs to subtract off the partition function for the vacuum solution to have a renormalized bound. Hence, we need to look at the difference of the two actions for the thermal AdS solution and the black hole solution. This is given by

\begin{equation}
-\beta(F_{BH}-F_{thermal})= \log Z_{BH} - \log Z_{thermal} = \left(\frac{Vol(S^{d-1})}{4 G_{N}}\right)\frac{-L^2_{AdS}r_{+}^{d-1}+r_{+}^{d+1}}{dr^2_{+}+ (d-2)L^2_{AdS}}
\end{equation}

\noindent
In the limit of high temperature, the black hole phase dominates, and one ends up with the following dependence for the logarithm of the partition function:

\begin{equation}
\log Z_{BH}  = \left(\frac{L^{d-1}_{AdS}}{4G^{(d+1)}_N}\right) \left(\frac{4\pi}{d}\right)^{d-1} \frac{2 \pi^{d/2}}{d \Gamma(\frac{d}{2})} T^{d-1} + \ldots
\end{equation}

\noindent 
One can see that for high temperatures, the mutual information varies with a lesser power of temperature than the logarithm of the partition function in keeping with the bound derived in the earlier section. For 2 dimensional CFTs, the difference of actions for the two phases is given by the following expression:

\begin{equation}
(\log Z_{BH} - \log Z_{AdS}) \big |_{d=2} = \left(\frac{L_{AdS}}{4 G_{N}}\right) \left(2\pi^2 L_{AdS} T - \frac{1}{2 L_{AdS} T} \right)
\end{equation}

\begin{figure}[h]
\centering
\begin{subfigure}[b]{.497\linewidth} \hspace{-0.4 cm} \vspace{-0.05 cm}
\includegraphics[width= \linewidth]{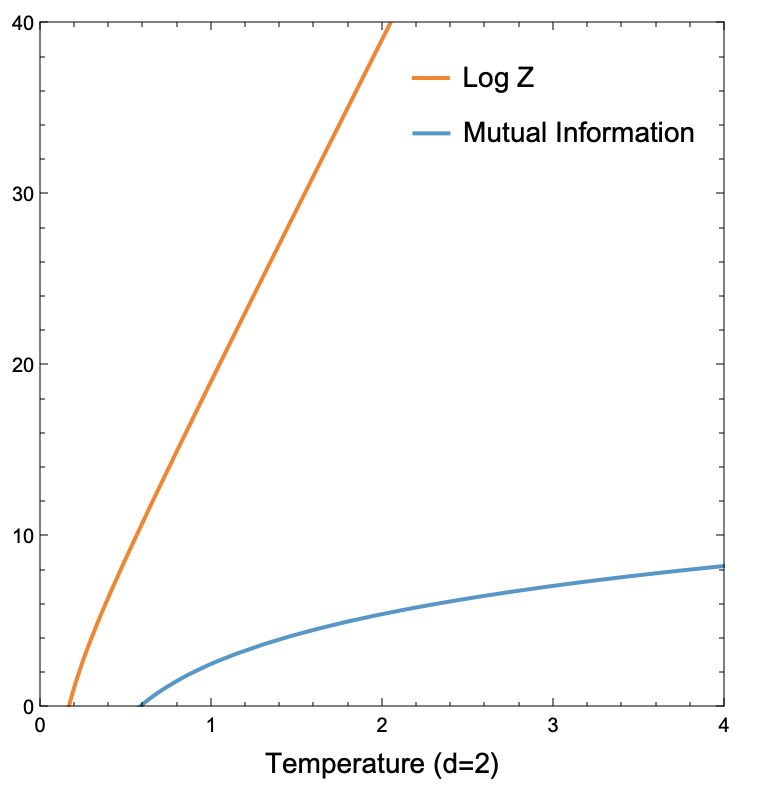}
\label{fig:1a}
\end{subfigure}
\begin{subfigure}[b]{.495\linewidth} \hspace{-0.50 cm} \vspace{-0.1 cm}
\includegraphics[width= \linewidth]{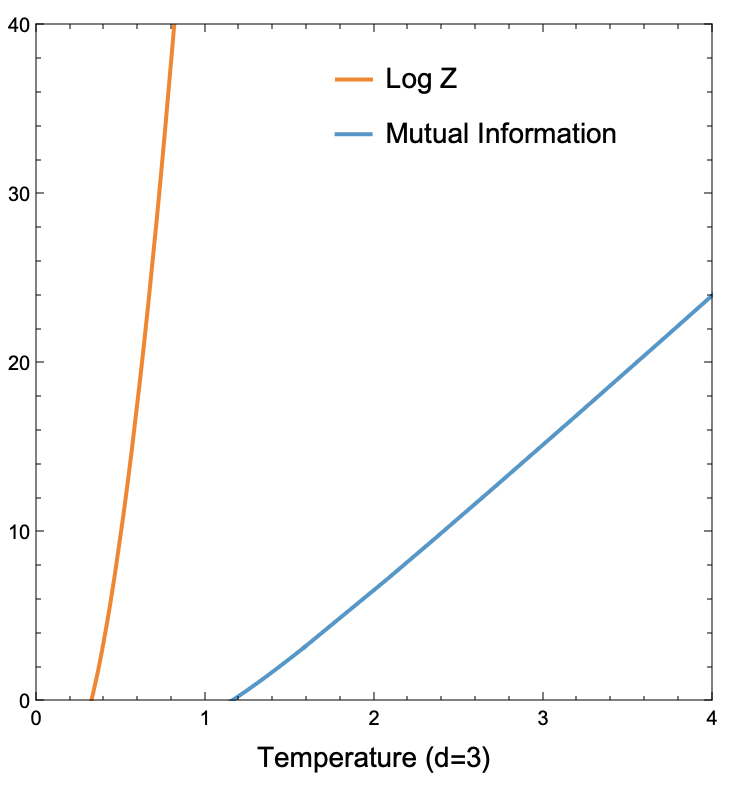}
\label{fig:2}
\end{subfigure}
\begin{subfigure}[b]{.49\linewidth} \hspace{-0.5cm}
\includegraphics[width= \linewidth]{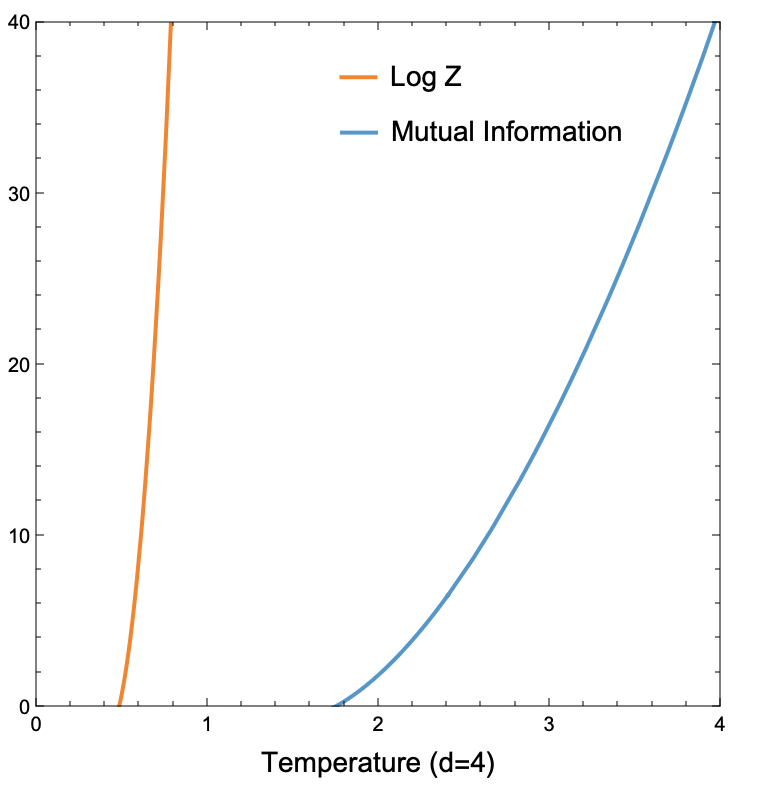}
\label{fig:3}
\end{subfigure}
\begin{subfigure}[b]{.48\linewidth} \hspace{-0.5cm} \vspace{0.04 cm}
\includegraphics[width= \linewidth]{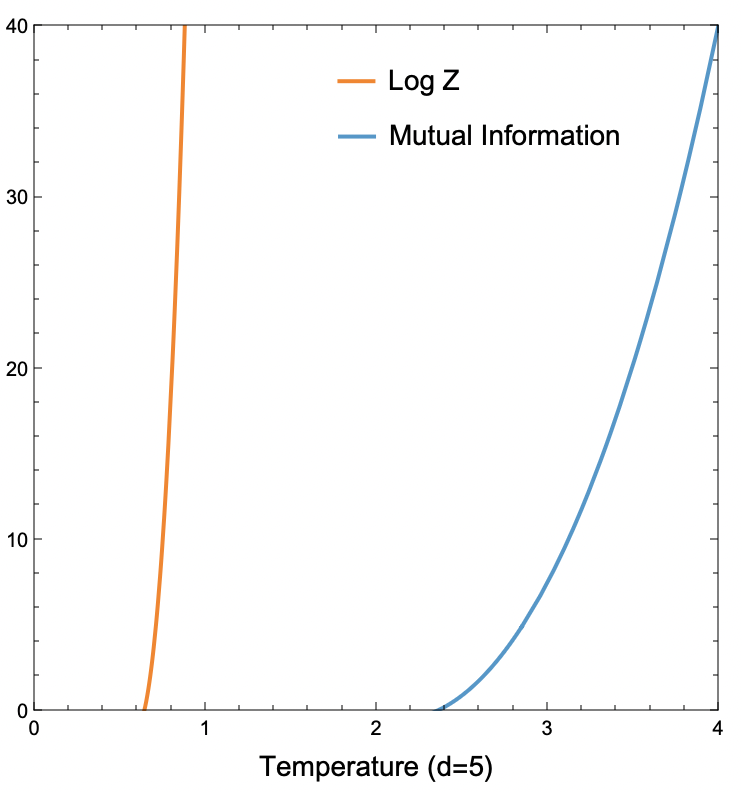}
\label{fig:4}
\end{subfigure}
\caption{Mutual Information and $\log Z(\beta)$ as a function of temperature for various dimensions $(d=2,3,4,5)$. Both quantities are zero before their respective critical temperatures $T_{HP}$ and $T_{MI}$. We set the factors $L_{AdS}$ and $4G_{N}$ to 1.}
\label{fig:1b}
\end{figure}

\section{From Nuclearity to Bulk Geometry}
\label{sect:9}
In this section, our aim is to relate the nuclear partition function constructed from modular theory to quantities that appear in the bulk geometry in the AdS spacetime for theories with holographic dual. In the previous section for the special case of diamonds in vacuum, we saw that the minimal bulk distance is related to the inverse temperature $\beta$ of the CFT partition function. In this section, 
we make some speculative proposals for the generalization of this result to more general regions and states.

We consider boundary regions with corresponding von Neumann algebras with $\mathcal{A} \subset \mathcal{B}'$ split. The dual description of $\mathcal{A}$ and $\mathcal{B}$ separately corresponds to disconnected entanglement wedges, and as usual there is a well-known correspondence between modular flow in the boundary and in the bulk \cite{Jafferis_2016,Faulkner_2017}. We will conjecture an additional aspect of this correspondence that allows one to approximately compute the modular nuclear partition function for these general regions. 

The expression we conjecture for the partition function involves a trace in a Hilbert subspace of heavy states. These are states generated by high conformal dimension single-trace scalar operators acting on vacuum (heavy operator limit in the bulk: $1 \ll m l_{AdS} \lesssim l_{AdS}/l_{string} \ll l_{AdS}/l_{Planck}$), where mass is taken to be large but not too large:

\begin{equation}
\begin{aligned}
Z(\beta, \Omega) &= \Tr_{\mathcal{H}} \left(\sqrt{\Delta^{-1/4}_{\mathcal{A}} \Delta^{1/2}_{\mathcal{B}'} \Delta^{-1/4}_{\mathcal{A}}}\right)\\
&\sim \quad  \sum_{i} \exp \left( - \int^{\tau_2}_{\tau_1} d \tau \sqrt{g_{\mu \nu} \dot{X}^{\mu} \dot{X}^{\nu}} \Delta_i - \int^{\tau_2}_{\tau_1} d \tau (\tilde{n} \cdot  \nabla n) J_i \right)
\end{aligned}
\end{equation}

\noindent 
We now explain the main aspects of this conjectured formula. Compared to the partition function in the symmetric case studied in the previous section (with $z= \bar{z}$), we replace the inverse temperature $\beta$ with the shortest distance between the disconnected bulk entanglement wedges of $\mathcal{A}$ and $\mathcal{B}$, the first integral in the exponent of (9.1). We also replace the angular velocity with a holonomy term, which computes the parallel transport of a specific time-like normal vector $n$ that is orthogonal to the tangent vector of the minimum geodesic. $\Delta_i$ and $J_i$ are boundary quantities corresponding to the conformal dimension and spin of heavy primary operators, where the trace is a sum over heavy states. In the bulk picture, the partition function describes the world-line action of a heavy spinning particle \cite{Castro_2014}, where the variable $\tau$ parametrizes the proper distance along the bulk geodesic. The worldline of a spinning particle becomes a \textit{ribbon}, where its twist is described by the second integral in (9.1). The covariant derivative is along the minimal bulk geodesic, where $\tau_1$ and $\tau_2$ are the end points of the geodesic. $\tilde{n}$ is the space-like normal vector to the geodesic between the two entanglement wedges. For sufficiently high temperatures, larger than order $1/\ell_{AdS}$ but lower than any $1/\ell_{Planck}$ dependent effects, the sum is dominated by states spanned by heavy operators acting on the CFT vacuum and our formula should apply.

We provide some evidence for (9.1) in lower dimensions. Choosing symmetric subregions, we can confirm this relation in $AdS_3/CFT_2$, details of which we relegate to Appendix \ref{sect:B}. Using the correspondence between the tips of causal diamonds and the conformal cross ratios from before, we can relate the bulk quantities to the corresponding boundary quantities $\beta$ and $\Omega$ that appear in the partition function in the following way:

\begin{equation}
\begin{aligned}
\beta &= \int^{\tau_2}_{\tau_1} d \tau \sqrt{g_{\mu \nu} \dot{X}^{\mu} \dot{X}^{\nu}}  = \cosh^{-1} \sqrt{z} + \cosh^{-1} \sqrt{\bar{z}} \\
\beta \Omega &= \int^{\tau_2}_{\tau_1} d \tau (\tilde{n} \cdot  \nabla n) = \cosh^{-1} \sqrt{z} - \cosh^{-1} \sqrt{\bar{z}}
\end{aligned}
\end{equation}

\noindent

For general dimensions we can write a slightly more detailed formula compared to (9.1) taking inspiration from the work of \cite{Chen_2018, Faulkner_2019}. In that work, an expression for modular flowed correlation functions in the bulk is given, where the geodesics are fine-tuned to cross the entanglement cuts. One of their key results was that the correlation functions involving $\Delta^{1/2}_{\Omega}$ between operator insertions calculated the length of a \textit{reflected geodesic}, which hits the entangling surface orthogonally before reflecting back. In the heavy operator limit, for which the geodesic approximation is valid, this gives rise to an expression that involves the mass of the heavy operator coupled with the geodesic length \cite{jafferis2014gravitydualsmodularhamiltonians}. Here, we wish to motivate a possible extension of that work involving higher spin operators, where in addition to the geodesic length, parallel transport of an orthonormal frame is also expected to play a role. That is, $\Delta^{1/2}_{\Omega}$ flips an orthonormal frame across the entangling cut, which becomes a time-reversal operation on the normal vector $n$. For a given modular operator corresponding to the bulk algebra $\mathcal{A}$ and a heavy operator contained in $\mathcal{A}$, we conjecture the following relation for the correlation functions:

\begin{equation}
\bra{\Omega} \mathcal{O}_{m_i,s_i} (x) \Delta^{1/2}_{\Omega;\mathcal{A}} \mathcal{O}_{m_i,s_i} (x) \ket{\Omega} \sim \exp\{-2m_i \ell{(x, X_\mathcal{A})}\} \exp\{-2 s_i \mathcal{S}(x, X_\mathcal{A}) \}
\end{equation}

\noindent
where $\ell{(x, X_\mathcal{A})}$ is the minimal geodesic between the location of the operator insertion and the extremal surface associated with algebra $\mathcal{A}$ in pure $AdS_{d+1}$ and $\mathcal{S}(x, X_\mathcal{A})$ is the holonomy term. This term is given by an $SO^{+}(1,d-1)$ transformation of an orthonormal frame along the geodesic. However, to relate this holonomy to the CFT partition function, where the conserved charges are given by the $SO(d-1)$ generators, and to the world-line action of the spinning particle in $AdS_{d+1}$, where the minimal geodesic parameterizes the proper time direction, one needs to perform a Wick rotation. Hence, the spin contribution to the two-point function in (9.3) becomes the following expectation value evaluated for wave functions labeled by irreducible representations of the group $SO(d)$:

\begin{equation}
\Longrightarrow \quad \left< U(g) \mathcal{T}_\mathcal{A} U(g^{-1}) \right>
\end{equation}

\noindent
where $U(g)$ implements the $SO(d)$ rotation in Euclidean time, and $\mathcal{T}_\mathcal{A}$ is a time-reversal operator acting on the normal frame at the entangling surface. We can use (9.3) to write an expression for the spin contribution of the partition function, which involves modular data for the two entangling cuts for the algebras $\mathcal{A}$ and $\mathcal{B}'$ and in this case involves a trace over the matrix representation. The matrix is comprised of a product of four matrices, implementing Euclidean rotation and time-reversal in an alternating way. Physically, this describes the parallel transport of a normal frame on the RT surface along the minimal geodesic and back (Figure 5). In two dimensions, this is expressed as follows.

\begin{equation}
\begin{aligned}
&\quad \quad \Tr \sqrt{R(i\theta) \mathcal{T}_{\mathcal{B}'} R(-i\theta) \mathcal{T}_{\mathcal{A}}^{-1}} \\
&= \quad
\Tr \sqrt{
\begin{pmatrix}
\cos (i \theta) & -\sin (i \theta)  \\
\sin (i \theta) & \cos (i \theta)
\end{pmatrix}
\begin{pmatrix}
-1 & 0 \\
0  & 1 
\end{pmatrix}
\begin{pmatrix}
\cos (i \theta) & \sin (i \theta) \\
-\sin (i \theta)  & \cos (i \theta)
\end{pmatrix}
\begin{pmatrix}
-1 & 0 \\
0  & 1 
\end{pmatrix} } \\
&= \quad \Tr (e^{- \theta \cdot J} )
\end{aligned}
\end{equation}

\noindent 
For higher dimensions, proper orthochronous Lorentz transformations describe the parallel transport of an orthonormal frame, where the generators can be written as a linear combination of a boost and a rotation generator. 

\begin{equation}
M(\alpha) \equiv e^{i \alpha_{\mu \nu} J^{\mu \nu}} = e^{i (\alpha_{i0} M^{i0} + \alpha_{ij} J^{ij})} \quad \mu, \nu = 0, \dots,d-1
\end{equation}

\noindent 
where $1 \leq i,j \leq d-1$ and $\alpha_{\mu \nu} \in \mathbb{R}$.
Performing a Wick rotation $t \rightarrow i \tau$, the boost becomes a complex Euclidean rotation, and the generator now is an element of the group $SO(d)$ instead of $SO^{+}(1,d-1)$. This can be expressed as follows.

\begin{equation}
 e^{i (\alpha_{i0} M^{i0} + \alpha_{ij} J^{ij})} \quad \rightarrow \quad e^{- \alpha_{i0} J^{i0} + i\alpha_{ij} J^{ij}}
\end{equation}

\noindent We can also use the well-known fact that proper
orthochronous Lorentz transformations, elements of the restricted Lorentz group $SO^{+}(1,d-1)$, can be written as a product of a pure boost and a rotation element. For our case, this amounts to showing that a Wick rotated element of the restricted Lorentz group has a unique polar decomposition $M= PU$, where $P$ is the positive self-adjoint element corresponding to the Euclidean time rotation and $U$ is the unitary transformation corresponding to the spatial rotations \cite{urbantke2002elementaryproofmorettispolar,jaffe}. This follows from the fact that the double cover of the restricted Lorentz group is $\text{Spin}(1,d-1)$, where every matrix $M \in \text{Spin}(1,d-1)$ has a unique polar decomposition, which also exists for the matrices $\Lambda(M) \in SO^+(1,d-1)$. 
Wick rotation leaves the unitary element unchanged, affecting only the pure boost. We find that the expression in (9.7) has the following polar decomposition:

\begin{equation}
e^{- \alpha_{i0} J^{i0} + i\alpha_{ij} J^{ij}} = e^{-\beta_{0i} J^{0i}} e^{i\beta_{ij} J^{ij}}
\end{equation}

\noindent where $\beta_{0i} , \beta_{ij} \in \mathbb{R}$. Putting this inside the trace, we can express the term associated with Berry holonomy:

\begin{equation}
\begin{aligned}
&\Tr\sqrt{ \left(e^{-\beta_{0 i} J^{0i}} e^{i \beta_{ij} J^{ij}} \right) \mathcal{T} \left(e^{- i \beta_{ij} J^{ij}} e^{\beta_{0 i} J^{0i}}\right) 
\mathcal{T}^{-1} } =\Tr \left(e^{- \beta_{0 i} J^{0i}} \right) \\ & \Rightarrow \quad \beta \Omega = \left( \sum_i \beta^2_{0 i} \right)^{1/2}
\end{aligned}
\end{equation}

\noindent
where all the generators are Hermitian with $\beta_{\tau i} , \beta_{ij} \in \mathbb{R}$. The above derivation tells us that regardless of how we boost or rotate to map from one diamond to the other, the two will be related by a general boost along some axis. The angular frequency is determined by the norm of the $d-1$ dimensional vector that is composed of the Euler angles for each $J^{0i}$ generator. This can be seen from the fact that spatial rotations are not affected by the time-reversal operator in the bulk. Putting everything together, we find the following expression for the conjectured full partition function in the heavy operator limit:

\begin{equation}
\Tr{\sqrt{\Delta^{-1/4}_{\Omega;\mathcal{A}} \Delta^{1/2}_{\Omega;\mathcal{B}'} \Delta^{-1/4}_{\Omega;\mathcal{A}}}} 
\thicksim \sum_{i,s} D_{i,s} \Tr_{s} \left(U(g) \mathcal{T}_{\mathcal{B}'} U(g^{-1}) \mathcal{T}_{\mathcal{A}}^{-1}\right)^{1/2} e^{-m_i l_s^{AB}} + O(e^{-2ml})
\end{equation}

\noindent 
where $D$ is the term associated with the degeneracy of the fields, $m_i$ is the mass of the quantum fields, the trace term is associated with their spin and $l_{AB}$ is the minimal geodesic between the two RT surfaces. It is also important to acknowledge that, in general, the geodesic couples to the spin of the particle and this could effect the details of our formula. (For a discussion on spinning geodesics, see \cite{Dyer_2017}.) Here, we ignore the spin contribution on the geodesic itself and write down the leading order classical answer, where mass and the spin contributions are decoupled. 

\section{Discussion}
\label{sect:10}

\noindent 
In this paper, we have shown that using the modular data of nested diamonds, one can assign a temperature to the mutual information phase transition in holography and that it happens at a temperature strictly higher than that of the Hawking-Page transition. We also provided two new bounds on mutual information in terms of the partition function, providing evidence for the order with which the two phase transitions occur. The temperature corresponding to the nested diamonds is found by looking at the minimum distance that separates them. 

We note that there is a natural generalization of this distance to non-AdS backgrounds. Assuming a semiclassical explanation of the Hagedorn divergence, we might expect that for any background, we can use these ideas to probe the minimal distance between the two disjoint entanglement wedges of the two boundary diamonds $\Diamond_1 \subset \Diamond'_2$. 

Then to find the distance in string units between the  wedges $\mathcal{E}(\Diamond_1)$ and $\mathcal{E}(\Diamond'_2)$, we can check 
for Hagadorn divergences in the bulk modular partition function\footnote{The use of this partition function should be contrasted with the boundary modular partition function, which will be swamped by the black hole transition. One could attempt to use this phase transition as a probe instead by replacing the critical temperature with the Hawking-Page temperature but this would be a much coarser, more inaccurate distance measure.}. In particular, we can
increase/decrease the effective temperature by considering instead the $p$-norm computed in the bulk theory. We conjecture the following expression for the distance between the entanglement wedges: 
\begin{equation}
d(\blacklozenge_1, \blacklozenge_2) = (\beta_{critical}) \sup \{p^{-1} : 0<p<\infty, \Vert \Delta^{-1/4}_{\mathcal{E}(\Diamond_1)} \Delta^{-1/4}_{\mathcal{E}(\Diamond_2)}  \Vert_{p,{\rm bulk}} < \infty \}
\end{equation}
We leave further exploration and of this formula to future work. 

\noindent


Another avenue for future work is to determine the temperature corresponding to nested diamonds for other backgrounds such as de Sitter space or other space times, which would then provide important data for a potential microscopic/holographic description of such space times.

The analytical expression for 
mutual information $I(\mathcal{A}: \mathcal{B})$ for holographic CFTs in higher dimensions is only known for symmetric intervals in equal time slices. This proves difficult to generalize to intervals at unequal times, where the spherical symmetry is no longer available (specifically, when $\mathcal{A}$ corresponds to an equal time interval and $\mathcal{B}$ is a boosted one). It is an interesting problem to determine the parameters $\beta$ and $\Omega$ that govern the phase transition of mutual information in such cases and to make comparisons with the Hawking-Page temperature in the AdS-Kerr background. We leave these for future work. 

During the completion of this work, we found that \cite{herderschee2025stringyalgebrasstretchedhorizons} appeared, where the authors study the connection between the minimum distance at which the split property breaks down or is restored for time-band algebras and nested causal diamonds, denoted by $\delta_s$ and the inverse Hagedorn temperature $\beta_H$ in holographic theories and provide upper and lower bounds for $\delta_s$ in terms of $\beta_H$. The authors also propose the Renyi reflected entropy $S^{n}_R(\mathcal{A}:\mathcal{B})$ as a diagnostic to determine the splitting distance $\delta_s$. It would be interesting to understand how our $d(\blacklozenge_1, \blacklozenge_2)$ relates to $\delta_s$ and the Renyi reflected entropy for the specific setups studied in \cite{herderschee2025stringyalgebrasstretchedhorizons} or whether it can be used to understand the quantum-connectedness of von Neumann algebras \cite{engelhardt2023algebraicereprcomplexitytransfer} and the emergence of type-III$_0$ algebras or other non-trivial structures when the split property breaks down. It would also be interesting and useful to provide upper or lower bounds for the Renyi reflected entropy in terms of the nuclear partition function studied in this paper. We leave these explorations for the future. 

\acknowledgments

This work is partially supported by the Air Force Office of Scientific Research under award number FA9550-19-1-0360 and the Department of Energy under award number DE-SC0015655.

\appendix

\section{Properties of Relative Entropy}
\label{sect:A}

In this section, we provide upper and lower bounds on the relative entropy between two normalized positive linear functionals denoted by $\mathcal{\sigma}$ and $\mathcal{\rho}= \sum_{i}^{n} p_i \rho_{i}$ belonging to a $C^*$ algebra with $p_i \geq 0$ and $\sum_i^n p_i =1$. The results here, which mainly follow \cite{petz}, are used to provide bounds on the mutual information for subregions in the CFT vacuum. The first property we wish to show is the following:

\begin{equation}
S(\rho, \sigma) = S(\sum_i^n \rho_i , \sigma ) \geq  \sum_{i}^n p_i S(\rho_i , \sigma) + \sum_i^n p_i \log p_i
\end{equation}

\noindent
where the second term is the negative of the Shannon entropy: $H(\{p_i\})= -\sum_i^n p_i \log p_i $

\noindent 
\textit{Proof:} This involves the use of Propositions 5.22 and 5.24 in \cite{petz}, the first of which is Donald's identity:

\begin{equation}
S(\sum_i^n p_i \rho_i) = \sum_i^n p_i S(\rho_i, \sigma) - \sum_i^n p_i S(p_i,\rho)
\end{equation}

\noindent
Following Ohya and Petz, we can prove this identity, using the language of conditional expectations. We can demonstrate the proof in n=2 and assume that $\sigma$ and $\rho$ are normalized states. Defining $\mathcal{N} = \mathcal{M} \oplus \mathcal{M}$. For elements $a,b \in \mathcal{M}$, we can define the conditional expectation $E: a \oplus b \mapsto \frac{1}{2}(a+b) \oplus \frac{1}{2}(a+b)$, which corresponds to the diagonal subalgebra $\mathcal{N}_0 = \{a \oplus a: a \in \mathcal{M} \}$. We also have the following linear functionals:

\begin{equation}
\begin{aligned}
\rho_{12}(a \oplus b) &= \rho_1(a) + \rho_2(b) \\
\sigma_{12}(a \oplus b) &= \frac{1}{2}(\sigma(a) + \sigma(b)) \\
(\rho_{12} \circ E) (a \oplus b) &= \frac{1}{2} (\rho_1(a) + \rho_2 (a) + \rho_1 (b) + \rho_2 (b))
\end{aligned}
\end{equation}

\noindent
Using the conditional expectation property of relative entropy suffices to prove Donald's identity:

\begin{equation}
\begin{aligned}
S(\rho_{12}, \sigma_{12} \circ E) &= S(\rho_{12}|_{\mathcal{N}_0}, \sigma_{12}|_{\mathcal{N}_0}) + S(\rho_{12}, \rho_{12} \circ E) \\
S(\rho_1, \frac{\sigma}{2})+ S(\rho_2,\frac{\sigma}{2}) &= S(\rho_1 + \rho_2, \sigma) + S(\rho_1,\frac{\rho_1+\rho_2}{2}) + S(\rho_2,\frac{\rho_1 + \rho_2}{2})
\end{aligned}
\end{equation}

\noindent
Having shown the proof for Donald's identity, we can now prove inequality (3.2) by focusing on the second term on the right-hand side:

\begin{equation}
S(\rho_i, \sum_i p_i \rho_i) \leq S(\rho_i, p_i \rho_i) = -\log p_i
\end{equation}

\noindent
Summing over $p_i$, one proves the inequality:

\begin{equation}
S(\rho, \sigma) \geq \sum_i^n p_i S(\rho_i, \sigma) - \sum_i^n p_i S(\rho_i, p_i \rho_i) = \sum_i^n p_i S(\rho_i, \sigma) +\sum_i^n p_i \log p_i
\end{equation}

\noindent
Here, we can also choose to write the relative entropy between the linear functionals in the following way:

\begin{equation}
\sum_i^n S(p_i \rho_i , \sigma) = \sum_i^n p_i ( S(\rho_i, \sigma) - S(\rho, p_i \rho_i) )
\end{equation}

\noindent
Having shown the lower bound on $S(\rho_i, \sigma)$, we now move onto the upper bound, which can be obtained simply by using Donald's identity or joint convexity. This yields

\begin{equation}
S(\rho, \sigma) \leq \sum_i^n p_i S(\rho_i, \sigma)
\end{equation}

\noindent
Combining everything, we have an upper and a lower bound:

\begin{equation}
\sum^n_i p_i S(\rho_i, \sigma) - H(\{p_i\}) \leq S(\rho,\sigma) \leq \sum^n_i p_i S(\rho_i, \sigma)
\end{equation}

\section{Calculations of bulk minimal geodesic and parallel transport of normal frame in \texorpdfstring{$AdS_3/CFT_2$}{AdS3/CFT2}}

\label{sect:B}

In this appendix, we follow the calculation of the minimal bulk geodesic and parallel transport of a normal frame between two extremal surfaces for the $AdS_3$ geometry, which are identified with the quantities that appear in the CFT partition function that is constructed from the $L_2$ nuclearity condition. We follow the derivation of \cite{Castro_2014} and \cite{Basu_2022}, where the same physics appears in slightly different contexts, where in the former, the authors compute the entanglement entropy in the presence of gravitational anomaly, where the entanglement functional describes the worldline of a spinning particle in the bulk. In the latter case, CFT quantities such as reflected entropy and entanglement negativity are computed from correlation functions of twist fields, where the geometric answer matches to what we have for the angular velocity term of the partition function. We follow the steps outlined in \cite{Basu_2022} to choose our disjoint boundary regions and parameterize the geodesic. We begin with a coordinate transformation on the $AdS_3$ metric:

\begin{equation}
ds^2 = \frac{-dt^2 + dz^2 + dx^2}{z^2} \quad \Longrightarrow \quad \frac{d\rho^2}{4\rho^2} + 2 \rho du dv
\end{equation}

\noindent
In the light-cone coordinates, the interval for the subregion $\mathcal{A}$ lies between the boundary points $(-u_1/2, -v_1/2)$ and $(u_1/2, v_1/2)$, while $\mathcal{B'}$ lies between $(-u_2/2, -v_2/2)$ and $(u_2/2, v_2/2)$. In $AdS_3$, the RT surface can be parametrized by a single parameter by solving the geodesic equation, which can be described by the following coordinates $(\rho(\lambda), u(\lambda), v(\lambda))$ \cite{Gao_2021}:

\begin{equation}
\begin{aligned}
u(\lambda) &= \frac{u_1}{2} \tanh \left( \lambda + \frac{1}{2} \log (u_1 v_1)  \right) \\
v(\lambda) &= \frac{v_1}{2} \tanh \left( \lambda + \frac{1}{2} \log (u_1 v_1)   \right)\\
\rho(\lambda) &= \frac{1}{2} \left( e^{\lambda} + \frac{e^{-\lambda}}{u_1 v_1} \right)^2 \\
\end{aligned}
\end{equation}

\noindent 
One of the normal vectors orthogonal to the minimal bulk distance $\beta$ is also the tangent vector to the entanglement wedge. The two normal vectors are given by the following:

\begin{equation}
\begin{aligned}
\dot{X}_{EW} &= \tilde{n}_1 = \left(\frac{du}{d\lambda}, \frac{dv}{d\lambda}, \frac{d\rho}{d\lambda} \right) \vline_{\lambda^*=-\frac{1}{2}\log(u_1v_1)} = \frac{u_1}{2} \partial_u + \frac{v_1}{2} \partial_v \\
n_1 &= -\frac{u_1}{2} \partial_u + \frac{v_1}{2} \partial_v \\
\dot{X}_{EWCS} &= \left(\frac{4i}{u_1 v_1}\right) \partial_\rho
\end{aligned}
\end{equation}

\noindent 
The orthonormal frame vectors orthogonal to the minimal geodesic (the same as the entanglement wedge cross section in $AdS_3$ for the commutant algebra) are chosen to satisfy the following constraints \cite{Castro_2014}:

\begin{equation}
\dot{X}_{EW}^2 = q^2 = -1 \quad, \quad \tilde{q}^2 = 1 \quad , \quad \dot{X}_{EWCS} \cdot q = \dot{X}_{EW} \cdot \tilde{q} = q \cdot \tilde {q} = 0 \quad, \quad \nabla q = \nabla \tilde{q} = 0
\end{equation}

\noindent Using these constraints, we have the following parallel transported orthonormal frame vectors $(q,\tilde{q})$:

\begin{equation}
\begin{aligned}
n(\tau) &= \cosh(\eta(\tau)) q(\tau) + \sinh(\eta(\tau)) \tilde{q}(\tau) \\
q &= z \partial_t = \frac{\sqrt{u_1v_1}}{2}(\partial_u +\partial_v) \\
\tilde{q} &= z \partial_x = \frac{\sqrt{u_1v_1}}{2}(\partial_u - \partial_v)
\end{aligned}
\end{equation}

\noindent
where $\tau$ parametrizes the minimal geodesic in the bulk. In terms of the endpoints of boundary intervals, the geodesic term and the transport term become the following:

\begin{equation}
\begin{aligned}
\beta &= \int^{\tau_2}_{\tau_1} d \tau \sqrt{g_{\mu \nu} \dot{X}^{\mu} \dot{X}^{\nu}} = \log \left(\frac{v_2 u_2}{v_1 u_1} \right)  \\
\beta \Omega &= \int^{\tau_2}_{\tau_1} d \tau (\tilde{n} \cdot  \nabla n) = \log \left(\frac{q(\tau_2) \cdot n_2 - \tilde{q}(\tau_2) \cdot n_2 }{q(\tau_1) \cdot n_1 -\tilde{q}(\tau_1) \cdot n_1 } \right) =  \log \left( \frac{v_1 u_2}{v_2 u_1} \right) 
\end{aligned}
\end{equation}

\noindent
where the arguments of the logarithms can be written in terms of the conformal cross-ratios. One simplification to note here is the fact that the boosted subregions are chosen to be symmetric around a midpoint to produce a minimal geodesic that is purely in the radial $z$ direction. Such a choice allowed us to keep track of the parallel transport of the orthonormal frame perpendicular to the bulk minimal geodesic, without affecting the geodesic itself with a simplified expression for the Berry holonomy, where the orthonormal frame undergoes time-reversal operation at the entangling surface. Completing the cycle with parallel transport in the reverse direction and another time reversal ensures that the gauge freedom to choose the orthonormal frame \cite{Castro_2014,Basu_2022} at the entangling cuts is canceled when we compute the holonomy.  

\begin{figure}[h]
  \centering
  \includegraphics[width=0.74\linewidth]{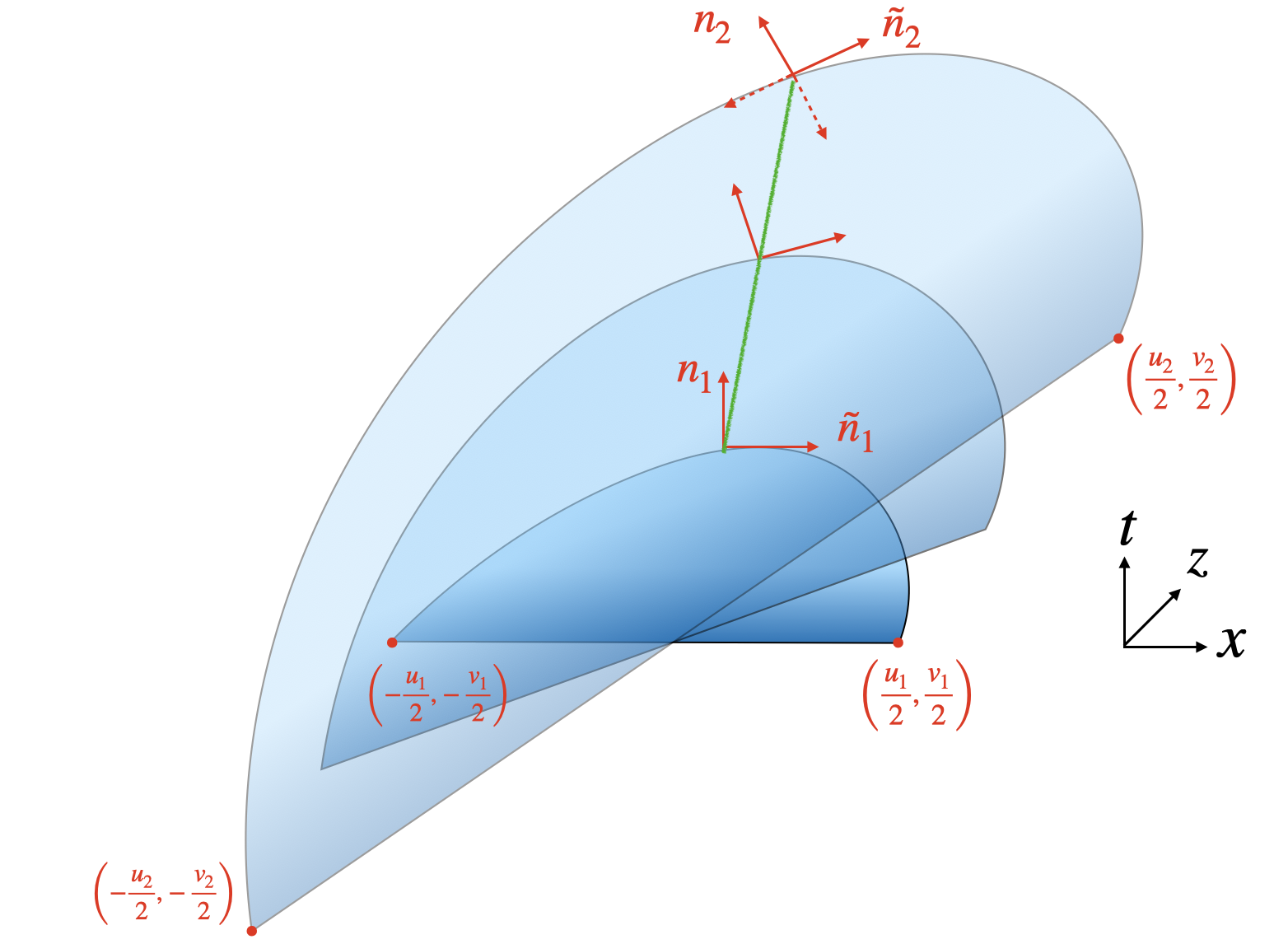}
  \label{fig:fig2a} 
  \caption{The parallel transport of an orthonormal frame along the minimal geodesic (green) between the entanglement wedges. The normal vectors are colored in red. Dashed vectors indicate time-reversed frame with respect to the entangling cut.}
\end{figure}
\noindent

\nocite{witten2022does}
\nocite{haag}
\nocite{dutta2019canonical}
\nocite{Levine_2021}
\nocite{Wen_2021}
\nocite{Wen_2022}
\nocite{Fonda_2015_2}
\nocite{Witten_2018}
\nocite{Asrat_2020}
\nocite{engelhardt2023algebraicereprcomplexitytransfer}
\nocite{gesteau2024stringyhorizons}
\nocite{liu2025lecturesentanglementvonneumann}




\bibliographystyle{plain} 
\bibliography{biblio} 

\end{document}